\documentclass[conference]{llncs}
\usepackage[letterpaper,hmargin=1.5in,vmargin=1.1in]{geometry}
\usepackage{subfigure}
\usepackage{dsfont,amsmath,amssymb,epsfig,theorem,url,multirow}
 \usepackage[colorlinks,
             bookmarks=false,
    pdfstartview={FitH}]{hyperref}

\newenvironment{XJenumerate}{
\begin{enumerate}
  \setlength{\itemsep}{1pt}
  \setlength{\parskip}{0pt}
  \setlength{\parsep}{0pt}}{\end{enumerate}
}

\begin{document}
\pagenumbering{arabic}

\title{Securing Interactive Sessions Using Mobile Device through Visual Channel and Visual Inspection}
 \author{
    Chengfang Fang \hspace{3.5cm}  Ee-Chien Chang
 }
 \institute{
School of Computing \\
National University of Singapore\\
    {\tt \{c.fang,
    changec\}@comp.nus.edu.sg}
 }
 
\maketitle
\begin{abstract}

Communication channel established from a display to a device's camera is known as \emph{visual channel}, and it is helpful in securing key exchange protocol~\cite{mccune2009seeing}.
In this paper, we study how visual channel can be exploited by a network terminal and mobile device to
jointly verify information in an interactive session, and how such information can be jointly presented in a user-friendly manner, taking into account that the mobile device can only capture and display a small
region, and the user may only want to authenticate selective regions-of-interests.
Motivated by applications in Kiosk computing and multi-factor authentication, we consider three security models: (1) the mobile device is trusted, (2) at most one of the terminal or the mobile device is dishonest, and (3) both the terminal and device are dishonest but they do not collude or communicate. We give two protocols and investigate them under the abovementioned models. We point out a form of replay attack that renders some other straightforward implementations cumbersome to use. To enhance user-friendliness, we propose a solution using {\em visual cues} embedded into the 2D barcodes and incorporate the framework of ``augmented reality'' for easy verifications through visual inspection. We give a proof-of-concept implementation to show that our scheme is feasible in practice.

\end{abstract}


\section{Introduction}
\label{sec:introduction}
Securing connection to a server through an untrusted network terminal is challenging even if the user has additional factor for authentication like one-time-password token, smartcard, or a mobile phone. One of the hurdles is the difficulty in securely passing information from the terminal to the device, and presenting the jointly verified authentic information to the user in a user friendly manner.
Using traditional channel to connect the device and  the terminal, like wireless connection or plug-and-play connection, are subjected to various man-in-the-middle attacks. Even if a secure channel can be established, it is still not clear how the additional device can help in authenticating subsequent messages rendered on the untrusted terminal's display.

\begin{figure}[!t] \centering
\subfigure[A bank transaction webpage.]{\includegraphics[width=0.32\textwidth]{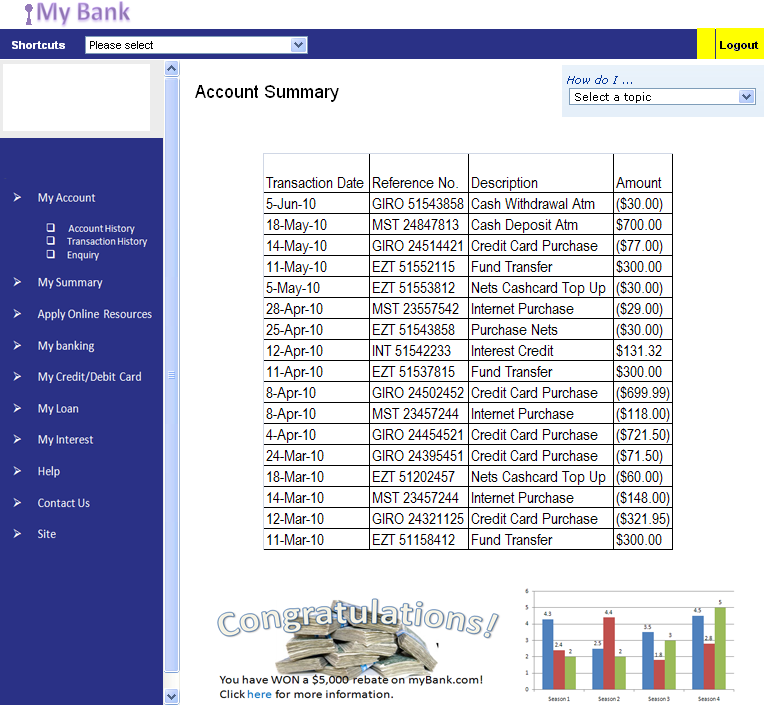}
\label{fig:OurScheme:origianltable}}
\subfigure[Method 1: mobile device is trusted.]{\includegraphics[width=0.32\textwidth]{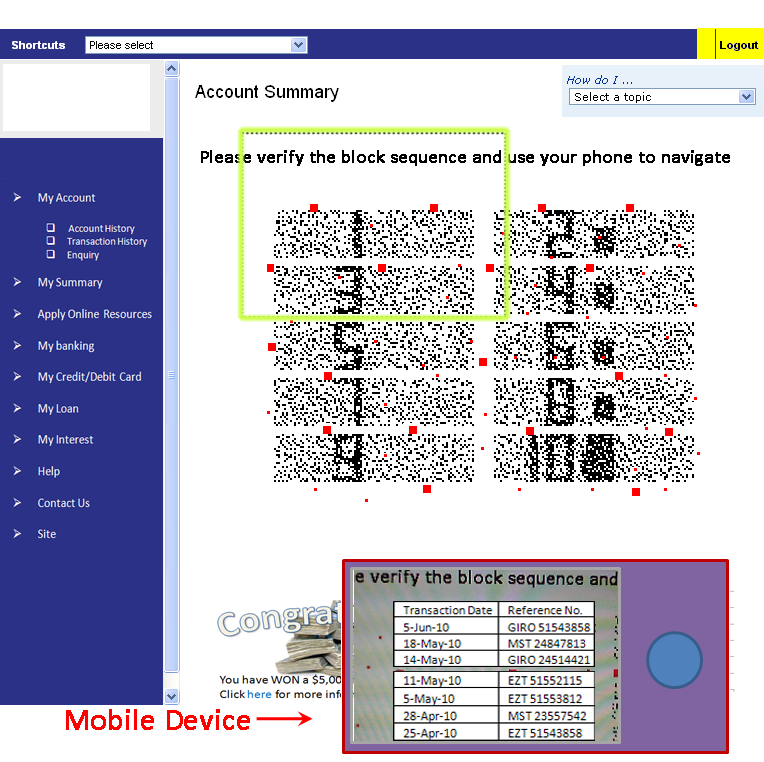}
\label{fig:OurScheme:MS1}}
\subfigure[Method 2: mobile device could be dishonest.]{\includegraphics[width=0.32\textwidth]{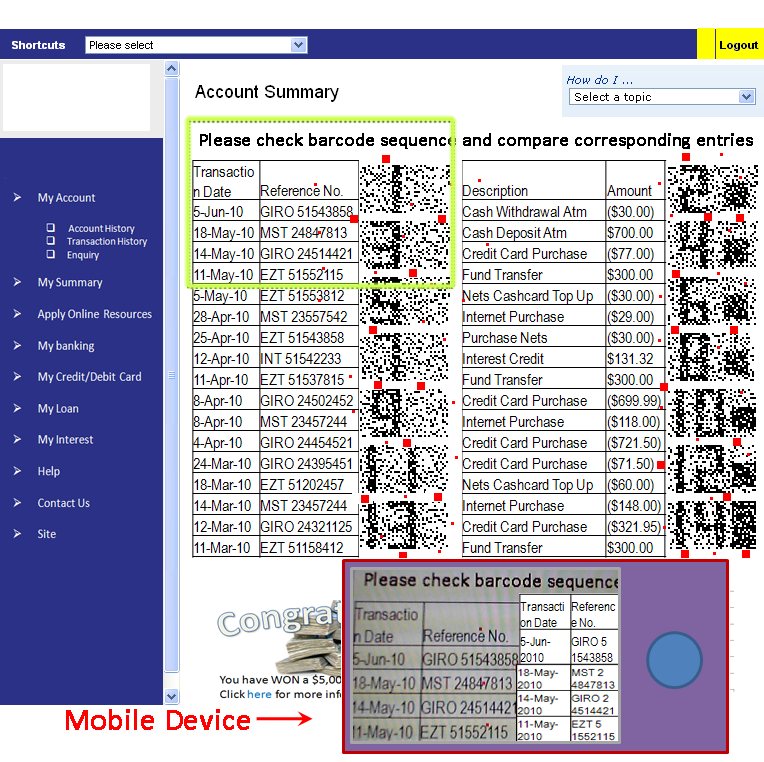}
\label{fig:OurScheme:MS23}}
\caption{Illustration of our schemes: Figure~\ref{fig:OurScheme:origianltable} is the bank transaction screenshot which contains a sensitive transaction table to be protected. Figure~\ref{fig:OurScheme:MS1} illustrates method 1 where the sensitive table is replaced by barcodes; and the mobile device captures, verifies and decodes part of the table. Figure~\ref{fig:OurScheme:MS23} illustrates method 2 where the sensitive table is displayed with barcodes; and the table is rendered on both terminal and mobile to be compared by the user. The decoded tables are generated by our proof-of-concept implementation which are then ``cut-and-paste'' to produce the illustration. The green boxes show the captured region and the red dots are for image registration.}
\label{fig:OurScheme}
\end{figure}

A number of recent works utilize cameras in the mobile devices to provide an alternative realtime communication channel from a display unit to a mobile device: messages are rendered  on the display unit in a form of, say 2D barcodes, which are then captured and decoded by the mobile device via its camera.  Although such visual channel could be eavesdropped by ``over-the-shoulder'' attacks, it is arguably impossible to modify or insert messages, and thus secure against man-in-the-middle attack.  Visual channel has been exploited in a few works in verifying the session key exchanged over an unsecured channel, for instance seeing-is-believing proposed by McCune et al.~\cite{mccune2009seeing}. There are also proposals on verifying untrusted display, for example, Clarke et al. propose verifying the display screen using stabilized camera device~\cite{clarke2002untrusted}.
In this paper, we  take a step further by investigating  authentication of  interactive sessions, with  consideration that many cameras  are
unable  to cover the whole screen in a single view with sufficient precision.  An example of interactive session is online banking application where a user can browse and selectively view pervious transactions, and carry out new transactions. A typical screenshot would contain important information like the user's account information, and less sensitive information like advertisements, help information, and navigation information, as shown in Figure~\ref{fig:OurScheme:origianltable}.

During an interaction session, after a session key  $k_s$ has been securely established between the server and the mobile device (could be established using seeing-is-believing~\cite{mccune2009seeing}), there could be many subsequent communication messages that require protection by $k_s$.
These messages may need to be rendered over different pages, or in a scrolling webpage where not all of them are visible at the same time. We remark that it is not clear how to protect them.
For instance, one may render the messages as 2D barcodes, each protected by the same $k_s$. To view  the message in a 2D barcode, the user moves the  mobile device over the barcode, and the device will capture, authenticate and display the message on its display panel.  However, as there are many barcodes associated with the same key, it is possible for a dishonest terminal to perform
``rearrangement'' attack: replays barcodes or shows barcodes in the wrong order.

The above attack arises due to the limitation that the camera is unable to capture the whole screen with sufficient precision, and not all messages can be rendered together in a single screenshot.  We treat the problem as the authentication of messages rendered in a sequence of large 2D regions, where only region in a small rectangular window can be captured at one time.
There are a few straightforward methods to overcome the rearrangement attack. For instance, one may prevent the attack by requiring the user to scan all the barcodes with his mobile device, and all the messages will be authenticated and rendered by the mobile device. However, it is troublesome for the user to scan all the barcodes, and there are situations where the user only wants to view some, but not all, of the messages. In addition, it is less preferred to navigate and browse the messages (e.g. a large table of transactions) within the relatively small display panel. In Section~\ref{sec:alternative},   we will discuss a few other straightforward methods and their limitations.

Our solution is to use a barcode scheme that given a message $m$ and a {\em visual cue} $v$,  is able to produce a barcode image that not only carries $m$ as its payload, but also visually appears as $v$ (see examples in Figure~\ref{fig:OurScheme:MS1} and Figure~\ref{fig:OurScheme:MS23}).  Our paper realizes such barcode scheme using technique borrowed from fragile image watermarking~\cite{cox1997secure}.  To embed a long messages into several barcodes, our main idea is to have a visual cue on each barcode indicating its position.
By visually inspecting the visual cues, the user can readily verify that the barcodes are in the correct arrangement. For example, in Figure~\ref{fig:OurScheme:MS1}, the visual cues are numeric numbers increasing by 1 from left to right, top to bottom. The black dot beside the number ``2'' indicates that the barcode is at the end of row, and the black block beside the number ``10'' indicates that it is the last (i.e. bottom-right) barcode. With the arrangement of barcodes verified, the user can then browse selective barcodes independently with his mobile device.

 In our security analysis, we consider the four parties setting where a user, who has a mobile device, wants to interact with a server via a network terminal. We focus on three security models. In the first model, the Internet terminal, including its CPU, keyboard and display unit, is untrusted by the user, whereas the mobile device is trusted. This model is motivated by the challenging problem in securing Kiosks~\cite{garriss2006towards,kauer2007oslo}, where Kiosks are untrusted public network terminal like workstations in Internet caf\'{e}.

In the second model, motivated by two-factor authentication, we consider scenarios where at least one of the terminal or mobile device is honest. We found that under the first model, it is possible to provide both confidentiality and authenticity; whereas under the second model, although authenticity can be achieved, it is not clear how to achieve confidentiality.

In the third model, we take one more step beyond two-factor authentication and consider a tricky setting where both the terminal and mobile device could be dishonest, but they do not collude in the sense that they do not know how to communicate with each other. This model is motivated by scenarios where the terminal and mobile device are compromised, but independently by two different adversaries, for instance, a dishonest mobile device that always says ``authentic'' for whatever authentication it is supposed to carry out,
and a network terminal that is tasked to deceive the user to accept a message given to the terminal. To detect such dishonest mobile device, our proposed method requires the mobile device to extract and produce a human readable proof from the authentication tag. A corresponding proof is also shown in the terminal's display and hence the user can visually verify whether they are consistent, as shown in Figure~\ref{fig:OurScheme:MS23}.

In addition to security requirements, user experience is also important.
Requiring the user to take snapshot of the screen is rather disruptive from the user's point of view.  We employ {\em augmented reality} to provide better user experience in verification. The design of our 2D barcode and the subregion authentication takes useability into consideration and fits nicely in the framework of augmented reality.
One example is as shown in Figure~\ref{fig:OurScheme:MS1}. The screenshot displayed by the terminal is a combination of sensitive data and non-sensitive data like advertisement and menu.  The sensitive data are replaced by 2D barcodes with visual cue as described before. The user treats the mobile device as an inspection device and places the mobile phone over the region to be inspected. In realtime, the mobile device captures and verifies the 2D barcode. If it is authentic, the decrypted message is displayed.  The non-sensitive portion of the screenshot is also displayed as it is to help the user to navigate.
We give a proof-of-concept system where we use a laptop equipped with a webcam to simulate the mobile device to show the feasibility of our methods.

\subsubsection*{Organization} \ \ \
We formally define our problem and three adversary models in Section~\ref{sec:definitionsandnotations}. Assuming the existence of a barcode scheme that is secure against rearrangement attack, we propose two protocols and analyze them under the three adversary models in Section~\ref{sec:protocol}. We give a construction for the required barcode scheme using visual cues in Section~\ref{sec:visualchannel} and discuss the design of visual cue symbols in Section~\ref{sec:subregionauthentication}. We compare our solutions with possible alternative methods in Section~\ref{sec:alternative}.  We describe our proof-of-concept implementation in Section~\ref{sec:implementation} and measure its performance in Section~\ref{sec:performance}. A discussion of existing work is given in Section~\ref{sec:relatedwork}. Section \ref{sec:conclusion} gives a conclusion of our paper.

\section{Models and Formulation}
\label{sec:definitionsandnotations}

There are four parties involved in our problem:  the user, the server, the mobile device and the network terminal. Let us call them {\tt User}, {\tt Server}, {\tt Mobile}, and {\tt Terminal}\ respectively.
In our framework, the term ``user'' literately refers to a person, and the mobile device is equipped with a camera, input device, a small display unit and a chip that can perform decoding of barcodes.

\label{ssec:problem formulation}

\begin{figure}[ht]
\centering
 \includegraphics[width=3.9in]{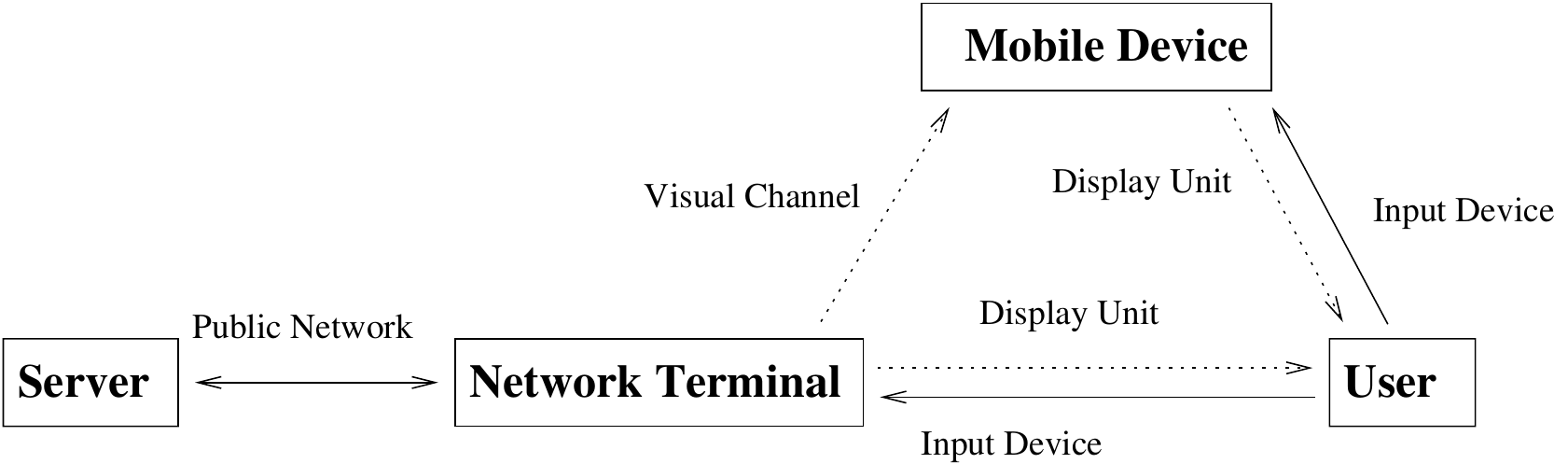}
  \caption{The communication channels among the four parties. }
 \label{fig:message passing}
\end{figure}

The communication channels among the four parties are as shown in Figure~\ref{fig:message passing}.
Note that there is no direct communication link between {\tt Mobile} and {\tt Server}. With 3G mobile network and WiFi connection widely available, one may argue that the model should consider such a link. Nevertheless, there are situations where the connection is not available due to cost or other constrains. In addition, there are also security concerns if the mobile device has Internet connection during the transactions: if {\tt Mobile} can directly send messages to {\tt Terminal}, they may collude and conduct coordinated attack.
Table~\ref{table:tableofnotation} gives a summary of our notations.

We consider the following security models for the channel between {\tt Server} and {\tt User}:

\begin{enumerate}
\item Model 1: {\tt Terminal} is not trusted by {\tt User},  but {\tt Mobile} is trusted and we want to protect both confidentiality and authenticity.
  \item Model 2: At least one of {\tt Terminal} and {\tt Mobile} is honest and we want to protect authenticity.
\item Model 3: Both {\tt Terminal} and {\tt Mobile} could be dishonest but they do not collude and we want to protect authenticity.
\end{enumerate}

In Model 3,
we treat the dishonest {\tt Terminal} and {\tt Mobile} as two different adversaries $A_{\tt T}$ and $A_{\tt M}$ with two different goals.
$A_{\tt T}$ is the dishonest terminal and its intension is to trick the user to believe that a given message $m'$ is authentic. The actual value of $m'$ is not determined prior to the connection. We can view it as a randomly chosen message that is passed to the $A_{\tt T}$. The adversary $A_{\tt M}$ is the dishonest mobile and has a easier goal: it is free to construct any message and trick the user to wrongly believe that it is authentic.  An example of $A_{\tt M}$ is one who always accepts whatever verification it is tasked to do. To capture the notion that they do not collude, we impose the restriction that $A_{\tt T}$  and $A_{\tt M}$ do not know how to communicate with each other, and the forge message $m'$ is randomly chosen and hold by one party. Hence, we exclude the attack where $A_{\tt T}$ covertly sends the message $m'$ to $A_{\tt M}$ through the visual channel.

\begin{table}
\centering \caption{Summary of Notations.}
\begin{tabular}{|l|l|}
\hline
$m_{\tt U}$& The message from {\tt User} to {\tt Server}.\\
$m_{\tt S}$& The message from {\tt Server} to {\tt User}.\\
$k_{\tt T}$& The key used in message authentication scheme.\\
$k_{\tt E}$& The key used in encryption.\\
$k_{\tt V}$& The key used in embedding visual cue.\\
$k_s$& The session key containing tuple $(k_{\tt T},k_{\tt E},k_{\tt V})$.\\
$v$& A visual cue symbol that carries the location information.\\
${\sf B}(k_s, m, v)$ & A barcode image encoding a message $m$ and visual cue $v$ with key $k_s$.\\
${\cal T}_{k_{\tt T}}(m)$ & An authentication tag of a message $m$ with key $k_{\tt T}$.\\
${\cal E}_{k_{\tt E}}(m)$ & An encryption of a message $m$ with key $k_{\tt E}$.\\
${\sf ECC}(m)$ & An error correcting encoding of a message $m$.\\
${\tt A} \rightarrow {\tt B} : m$ & The entity {\tt A} sends a message $m$ to another entity {\tt B}.\\
${\tt A} \stackrel{{\tt C}}{\longrightarrow} {\tt B} : m$ & The entity {\tt A} sends a message $m$ to {\tt B} using {\tt C} as a relay point.\\
\hline
\end{tabular}
\label{table:tableofnotation}
\end{table}

\section{Protocols}
\label{sec:protocol}
We now give our proposed protocols for securing the communication between {\tt Server} and {\tt User} assuming we have a barcode embedding technique that can protect the integrity and confidentiality of its payload, and visible visual cue can be rendered onto the barcode to indicate the barcode location as in Figure~\ref{fig:OurScheme:MS1}. Given a message $m$, a visual cue $v$, and a session key $k_s$, let us write the barcode (represented as images) as ${\sf B}(k_s, m, v)$.
For clarity in presentation, we first consider the case where the message can be embedded into one barcode block whose size is small enough to be entirely captured by {\tt Mobile}'s camera with sufficient precision. Thus, we take the visual cue as a single dot, indicating to the user that there is only a single barcode to be read. We will later study the case for multiple messages in Section~\ref{sec:visualchannel} and Section~\ref{sec:subregionauthentication}.

We assume that {\tt Server} has already established a long term shared key with {\tt Mobile} when the user registers an account with the server.
In additional, for model 2 and 3, we assume that {\tt User} has established a password with {\tt Server}.
Before each interactive session, {\tt Server} authenticates {\tt User} and {\tt Mobile} to get a session key $k_s$.
A secure key exchange can be derived from modified seeing-and-believing~\cite{mccune2009seeing} and combination of the proposed method in this section. Due to space constrain, we do not include details in this paper.

\subsection{Server to User}
\label{ssec:protocolmodel2}

Consider the case where {\tt Server} wants to send a message $m_{\tt S}$ to {\tt User}. We propose two methods, denoted MS1 and MS2 (message from server), where method MS1 is more user-friendly compared to MS2, but it requires that {\tt Mobile} is trusted.

\subsubsection*{MS1.} \ \ \
To send a message $m_{\tt S}$ to {\tt User}, the following steps are carried out.
(1) {\tt Server} generates a barcode image ${\sf B}(k_s, m_{\tt S}, v)$ and sends the barcode to {\tt Terminal}.
 (2) {\tt Terminal} displays the barcode.
 (3)  {\tt User} inspects and verifies the visual cue is valid.
(4) {\tt Mobile} captures the barcode and rejects if the barcode is not authentic.
(5) {\tt Mobile} renders $m_{\tt S}$ on its display.
(5) {\tt User} reads and accepts $m_{\tt S}$ from {\tt Mobile}'s display panel.
 Below is a summary for MS1:
\ \\

\hrule
\begin{enumerate}
\item ${\tt Server} \rightarrow{\tt Terminal}$: ${\sf B}(k_s, m_{\tt S}, v)$;
\item ${\tt Terminal} \rightarrow {\tt User}$: $v$;
\item  {\tt User} verifies $v$;
\item ${\tt Terminal} \rightarrow {\tt Mobile}$: ${\sf B}(k_s, m_{\tt S}, v)$;
\item ${\tt Mobile} \rightarrow {\tt User}$: $m_{\tt S}$;
\item ${\tt User}$ accepts $m_{\tt S}$.
\end{enumerate}
\hrule

\subsubsection*{MS2.} \ \ \
The main difference in this method from the previous MS1 is that, the message $m_{\tt S}$ is displayed by both {\tt Terminal} and {\tt Mobile} for {\tt User} to verify, and thus {\tt User} is able to detect if one of them is dishonest.
 (1)   {\tt Server} first generates a barcode image ${\sf B}(k_s, m_{\tt S}, v)$, then it sends both the barcode image and the message $m_{\tt S}$ to {\tt Terminal}.
(2) {\tt Terminal} displays the barcode, side-by-side with $m_{\tt S}$.
 (3)  {\tt User} inspects and verifies the visual cue is valid.
(4) {\tt Mobile} captures the barcode and rejects if the barcode is not authentic, otherwise, displays $m_{\tt S}$.
(5) {\tt User} reads $m_{\tt S}$ from {\tt Mobile}'s display panel and {\tt Terminal}'s display.
(6) {\tt User} accepts $m_{\tt S}$ if the $m_{\tt S}$ in step (2) is consistent with $m_{\tt S}$ in step (4).
 Below is a summary for MS2:

\ \\

\hrule
\begin{XJenumerate}
\item ${\tt Server} \rightarrow{\tt Terminal}$: ${\sf B}(k_s, m_{\tt S}, v)$, $m_{\tt S}$;
\item ${\tt Terminal} \rightarrow {\tt User}$: $v$, $m_{\tt S1}$;
\item  {\tt User} verifies $v$;
\item ${\tt Terminal} \rightarrow {\tt Mobile}$: ${\sf B}(k_s, m_{\tt S}, v)$;
\item ${\tt Mobile} \rightarrow {\tt User}$: $m_{\tt S2}$;
\item ${\tt User}$ accepts $m_{\tt S1}$ if $m_{\tt S1} = m_{\tt S2}$.
\end{XJenumerate}
\hrule

\subsection{User to Server}
\label{ssec:protocolmodel1}

Now we consider the following methods MU1 and MU2 (message from user) for sending the message $m_{\tt U}$ to {\tt Server}. Method MU1 protects both confidentiality and authenticity of the message, whereas method MU2 protects only the authenticity but involves less user operation.

\subsubsection*{MU1.} \ \ \
MU1 consist of the following steps to send a message $m_{\tt U}$ to {\tt Server}.
(1) {\tt User} enters $m_{\tt U}$ to {\tt Mobile}.
(2) {\tt Mobile} computes and shows {\tt User} the encrypted form ${\cal E}_{k_{\tt E}}(m_{\tt U})\|{\cal T}_{k_{\tt T}}({\cal E}_{k_{\tt E}}(m_{\tt U}))$ in readable characters (for e.g. using {\tt uuencode}).
(3) {\tt User} sends displayed string to {\tt Server} through {\tt Terminal}'s input device.
(4) {\tt Server} accepts $m_{\tt U}$ if the tag is valid.  Below is a summary for MU1: \\

\hrule
\begin{XJenumerate}
\item ${\tt User} \rightarrow {\tt Mobile}$: $m_{\tt U}$;
\item ${\tt Mobile}\rightarrow {\tt User}$: ${\cal E}_{k_{\tt E}}(m_{\tt U})\|{\cal T}_{k_{\tt T}}({\cal E}_{k_{\tt E}}(m_{\tt U}))$;
\item ${\tt User} \stackrel{{_{\tt Terminal}}}{\longrightarrow} {\tt Server}$: ${\cal E}_{k_{\tt E}}(m_{\tt U})\|{\cal T}_{k_{\tt T}}({\cal E}_{k_{\tt E}}(m_{\tt U}))$;
\item {\tt Server} accepts $m_{\tt U}$ if the tag ${\cal T}_{k_{\tt T}}({\cal E}_{k_{\tt E}}(m_{\tt U}))$ is valid.
\end{XJenumerate}
\hrule

\subsubsection*{MU2.} \ \ \
In scenarios where the confidentiality of $m_{\tt U}$ is not required, we can employ a more user friendly protocol MU2 as follow:
(1) {\tt User} enters $m_{\tt U}$ through {\tt Terminal}'s input device, and {\tt Terminal} forwards $m_{\tt U}$ to {\tt Server}.
(2) {\tt Server} generates a barcode  ${\sf B}(k_s, m_{\tt U}\|c, v)$, where $c$ is a randomly generated nonce and $\|$ means concatenation. {\tt Server} sends the barcode to {\tt Terminal}.
(3) {\tt Terminal} displays the barcode, and {\tt User} visually verifies that the visual cue $v$ is correct.
(4) {\tt Mobile} captures the barcode and rejects if the barcode is not authentic.
(5) {\tt Mobile} renders the message $m_{\tt U}$ and the nonce $c$ on its display.
(6) If $m_{\tt U}$ is consistent with the message {\tt User} entered in step (1), {\tt User} enters $c$ to {\tt Terminal}, and {\tt Terminal} forwards it to {\tt Server}.
(7) {\tt Server} rejects if the nonce $c$ is wrong.

Although involves more steps,  MU2 is less tedious from the {\tt User}'s point of view, since {\tt User} does not need to enter $m_{\tt U}$ using {\tt Mobile}'s input device.  The corresponding steps for MU2 are summarized below:
\ \\

\hrule
\begin{XJenumerate}
\item ${\tt User} \stackrel{{_{\tt Terminal}}}{\longrightarrow} {\tt Server}$: $m_{\tt U}$;
\item ${\tt Server} \rightarrow {\tt Terminal}$ : ${\sf B}(k_s, m_{\tt U}\|c, v)$;
\item ${\tt Terminal} \rightarrow {\tt User}$: $v$;
\item ${\tt Terminal} \rightarrow {\tt Mobile}$: ${\sf B}(k_s, m_{\tt U}\|c, v)$;
\item ${\tt Mobile} \rightarrow {\tt User}$: $m_{\tt U}, c$;
\item ${\tt User} \stackrel{{_{\tt Terminal}}}{\longrightarrow} {\tt Server}$: $c$;
\item {\tt Server} accepts $m_{\tt U}$ if $c$ is consistent, rejects otherwise.
\end{XJenumerate}
\hrule

\subsection{Analysis}
\label{sec:analysisofprotocols}

In this section, we analyze our methods under different adversary models.

\subsubsection*{Model 1 ({\tt Mobile} is trusted)} \ \ \
In Model 1, we use MU1 for sending message to {\tt Server}, and use MS1 for {\tt Server} to send message to {\tt User} to achieve confidentiality and authenticity of the communication channel.

For both methods, {\tt Terminal} plays the role of a relay point for passing message and thus a malicious {\tt Terminal} is the man-in-the-middle.
Hence, this is the classical setting where the two end points ({\tt Server} and {\tt Mobile}) having a shared key want to communicate over a public channel.  The cryptographic technique (encryption and message authentication code) can secure the channel and provide both confidentiality and authenticity.

It is clear that MU2 and MS2 cannot protect the confidentiality under this model as the messages are sent in clear through {\tt Terminal}, and thus they are not suitable in this model.

\subsubsection*{Model 2 (At least one is honest)} \ \ \ In Model 2,
we use MU2 to send message to {\tt Server}, and use MS2 for {\tt Server} to send message to {\tt User}. We want to achieve authenticity of the message $m_{\tt S}$. We are not interested in confidentiality here. It is an interesting future work to investigate whether confidentiality can be achieved under this model. Since we are not sure which of {\tt Mobile} and {\tt Terminal} is dishonest, it is not clear whether confidentiality can be achieved.

Suppose {\tt Terminal} is dishonest. In both directions of the communication, we can treat the barcode as the MAC of the message, $m_{\tt U}$ and $m_{\tt S}$ respectively, and {\tt Terminal} does not have the key.  Similar to analysis for Model 1, this is a classical setting and the authenticity of the message inherit from the MAC we used in the barcode construction.

On the other hand, let us consider the case where the {\tt Mobile} is dishonest.
In MU2, {\tt Terminal} is honest and will forward $m_{\tt U}$ to {\tt Server} as it is, thus, it is impossible for {\tt Mobile} to modify $m_{\tt U}$ without {\tt Server} notices. Similarly, in MS 2, since the actual message $m_{\tt S}$ is displayed by the honest {\tt Terminal}, {\tt User} can compare the displayed message and thus any modification can be detected.

Note that MU1 and MS1 is not secure in this model: if {\tt Mobile} is dishonest and change the message to $m'$, there is no way for {\tt User} or {\tt Server} to verify it.

\subsubsection*{Model 3 (No collusion)} \ \ \
It turns out that the protocol we used in method 2, i.e. MU2 and MS2, can achieve authenticity in this model as well.

Let us first analyze MU2. Recall that the goal of a dishonest {\tt Terminal} is to trick {\tt Server} to accept a message $m_{\tt U}'$. To do so {\tt Terminal} must send {\tt Server} the message $m_{\tt U}'$, and obtain a barcode $b$ contains $m_{\tt U}'$ and $c$. {\tt Server} accepts $m_{\tt U}'$ only if the verification code $c$ is presented. Since $c$ is randomly chosen,  {\tt Terminal} is unlikely to succeed in guessing $c$. Therefore, he needs to get $c$ from user.  Without any hint from {\tt Terminal}, {\tt Mobile} is not able to display the message that the user is expecting.

Now let us analyze MS 2. In this case the dishonest {\tt Terminal} wants to trick {\tt User} into accepting a message $m_{\tt S}'$. To achieve the goal, it must display $m_{\tt S}'$ side-by-side with the barcode. As {\tt Terminal} does not know the key $k_s$ he is unable to forge the barcode. Now, consider the dishonest {\tt Mobile}. Recall that there is no communication from the {\tt Terminal} to {\tt Mobile}, the {\tt Mobile} is unable to display the message $m_{\tt S}'$ which is required to trick {\tt User} to accept $m_{\tt S}'$.

\begin{table*}[ht]
\centering \caption{Summary of Methods.}
\begin{tabular}{|c|c|c|c|c|c|}
\hline
& MU 1 & MU 2 & MS 1 & MS 2 \\
\hline
Model 1                 & C, A, U1 & A, U1, U2    &C, A, U1, U2&  A, U1, U2    \\
Model 2                 & N     & A, U1, U2    &N      & A, U2\\
Model 3    & N    & A, U1, U2    &N     & A, U2 \\
\hline
\end{tabular}
\label{table:summaryofmethod}
\scriptsize{ \ \\ \ \\ {\bf Note:} C, A, N are related to security goals and U1, U2 are related to useability.\\
 C: confidentiality is achieved; A: authenticity is achieved; N: none of C and A can be achieved.\\
 U1: no user comparison of messages is required; U2: no user input via {\tt Mobile}'s input device is required.}
\end{table*}

Table~\ref{table:summaryofmethod} summarizes the security and user friendliness of our methods under different models.

\section{Visual Channel}
\label{sec:visualchannel}

A main component in building our visual channel is the construction of 2D barcode with visual cues: given a secret key $k_s = (k_{\tt T},k_{\tt E},k_{\tt V})$, a message $m$, and a visual cue symbol $v$ we want to produce a 2D barcode ${\sf B}(k_s, m, v)$ such that the cue $v$ is clearly visible, and the message $m$ can be extracted under noise. On the other hand, there are security requirements on the confidentiality of $m$ and integrity of $m$ and $v$. Any modification on $m$ and $v$  must be detected.

\subsection{Construction Overview}

There are a number of stages of the visual channel construction:
\begin{XJenumerate}
\item (Encryption-then-MAC): Given $m$, and the keys $k_{\tt E},k_{\tt T}$, the message $m$ is protected using encryption and MAC with key $k_{\tt E}$ and $k_{\tt T}$ respectively, and get $m_0$ = ${\cal E}_{k_{\tt E}}(m_{\tt U})\|{\cal T}_{k_{\tt T}}({\cal E}_{k_{\tt E}}(m_{\tt U}))$.
\item (Error correcting): Error correcting code is then applied on the result $m_0$, and get ${\sf ECC}(m_0$), let us call this $m_1$.
\item(Embedding visual cue): Given a message $m_1$,  a key $k_{\tt V}$, and a visual cue $v$ represented as a 2D array of bits, the $m_1$ is embedded into a larger 2D array of bits $I$ which visually appear as $v$, Section~\ref{ssec:encodinganddecoding} gives details on the embedding process.
\item (Adding control point and rendering): A set of control points(red dots in Figure~\ref{fig:OurScheme:MS1}) is then added around $I$ for image registration purpose.
\end{XJenumerate}

Thus, our barcode is a black and white image with red pixels.

\subsection{Encoding with Visual Cue}
\label{ssec:encodinganddecoding}

When a message is too large, multiple barcodes are required to encode it. As mentioned in the introduction, multiple barcodes protected by a single session key are subjected to  ``rearrangement'' attack. To detect the attack, we propose binding  location information to the barcode using visual cue.  This section gives a method in embedding the visual cue. Note that the process of embedding  a visual cue to a barcode is essentially  digital watermarking, where the visual cue is the host, and the barcode is a message to be ``watermarked'' to the host.

\begin{figure}[ht] \centering
\subfigure[Two groups of L-blocks.]{
\includegraphics[width=2.5in]{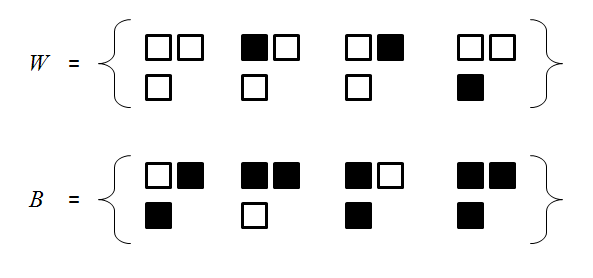}
 \label{fig:lgroup}}
\subfigure[Tile up with L-blocks.]{
\includegraphics[width=2in]{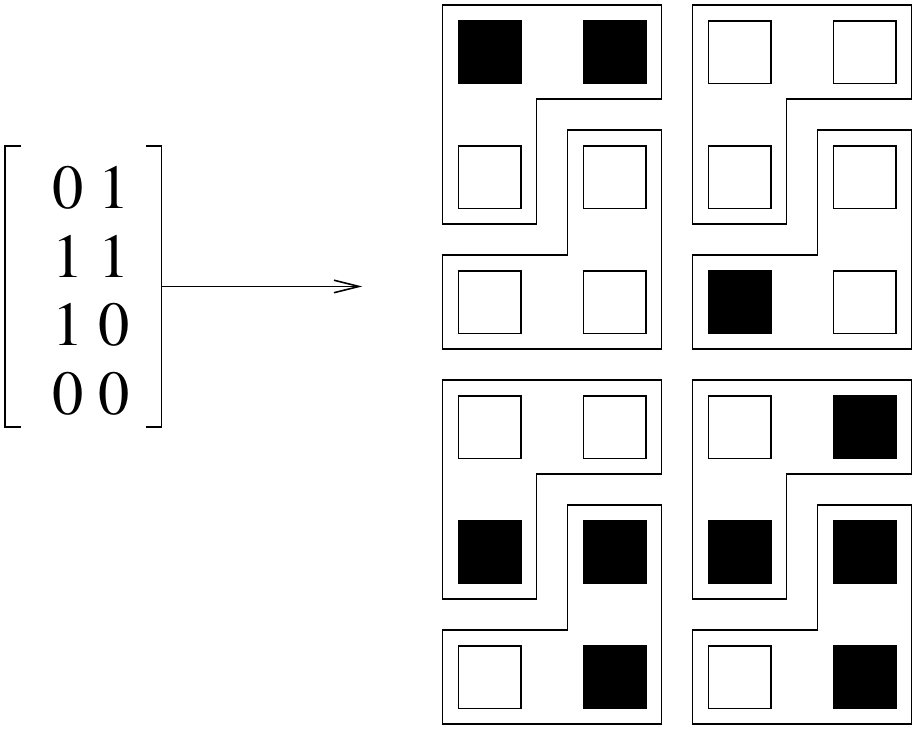}
 \label{fig:lshape}}
\caption{L-blocks for constructing visual cues}
\end{figure}

Given a $n$-bits message $m_1$, let us arrange it as a $x$ by $y$ binary matrix where  $n=x\cdot y$ and $x$ is even. Let us assume that the given visual cue is a $x/2$ by $y$ pixels image where each pixel is either 0 (representing a black pixel) or 1 (representing a white pixel).
Therefore, every  $2$ bits in $m$ is associated with $1$ pixel of the visual cue, and together they can be represented with  $3$ black-and-white pixels in the final barcode. The $3$ pixels are arranged in a ``L''-shape as
shown in Figure~\ref{fig:lgroup}. Let us call the $3$ pixels as
a L-block. The $2^3$ combination of values in a L-block is divided
into two groups: $W$ and $B$. The L-blocks in $W$ have more white
pixels and thus the L-blocks appear as
``white''. Conversely, the L-blocks in $B$ will appear as ``black''.

 Given a binary value
$v_1 \in \{0,1\}$ of a pixel of the visual cue image, we want to encode two bits $\langle b_1, b_2\rangle$  into a three pixels
L-block, such that the brightness of the L-block can be adjusted according to $v_1$. For instance, if $v_1 = 1$, the encoding outputs
only elements in $W$. Since there are $4$ elements in $W$, it is
possible to encode the two bits $b_1$ and $b_2$.
Beside for the value of $v_1$, there is no further constraint on how the encoding
of $\langle b_1, b_2\rangle$ to the $4$ elements in $W$ is to be done. In order to prevent the adversary from modifying the appearance of the visual cue, the mapping from the $2$ bits $\langle b_1, b_2\rangle$ to the three pixels of the associated L-block, $\langle p_1,p_2,p_3\rangle$, has to be kept secret. Hence, the key space for encoding a bit pair is $4!
\times 4! = 576$.

To decode a barcode, {\tt Mobile} applies the decoding and decryption functions in a reverse order and ignore the bit $v_1$. That is, it first extracts the bit pairs from every L-blocks, and get the message $m'$. Next, error correcting is applied and the authenticity of the message can be verified.

\subsection{Security Analysis}

We would like our barcode scheme to achieve the following properties: (1)authenticity and confidentiality of $m_{\tt S}$ and (2) the integrity of visual cue.

\subsubsection*{Authenticity and confidentiality of message} \ \ \
The authenticity and confidentiality of the message embedded in our barcode scheme rely on the security of the underlying encryption and message authentication scheme. Bellare et al.~\cite{bellare2008authenticated} show that when the encryption ${\cal E}_{k_{\tt E}}$ achieve indistinguishability under chosen-plaintext attack (IND-CPA), and the message authentication scheme ${\cal T}_{k_{\tt T}}$ is strongly unforgeable (SUF-CMA), then the Encrypt-then-MAC composition method achieves IND-CPA, INT-CTXT (integrity of ciphertexts) and IND-CCA ((adaptive) chosen ciphertext attack).

\subsubsection*{Integrity of visual cue} \ \ \
An adversary may try to modify some L-shape
blocks such that the visual cue on two barcode blocks are swapped, and thus, he can rearrange the two blocks without being detected.
As discussed in Section~\ref{ssec:encodinganddecoding}, any
modification of an $L$-shape block's brightness will have $\frac{1}{4}$ chance of not being detected. Suppose at least $\beta$
number of $L$-shape blocks have to be modified in order to deceive
the user, then the chances of not being detected will be
$(\frac{1}{4})^\beta$, where $\beta$ depends on the size of a barcode block, and the visual cue design.

However, the above analysis does not hold when we consider the whole process
of decoding, where the error correction is included. Recall that, due to inevitable noise, we need to apply error correcting before extracting ${\cal E}_{k_{\tt E}}(m_{\tt S})$. Therefore, when small number of $L$-shape blocks
are corrupted, the payload $m_1$ can still be correctly decoded.
Hence, the choice of error-correction and the
design of the cues cannot be done separately. Furthermore, some
error-correction code can correct more errors than its guaranteed
level in some situations. Due to the concern of forgery, it is
important not to correct those errors.

To prevent an adversary from making small changes that can deceive the user and yet get verified, one design consideration of the visual cue is to choose symbols with large mutual Hamming distance from each other. In our implementation to be described in Section~\ref{sec:implementation}, we use numerical digits as visual cue, where the minimum hamming distance for two symbols is $14$ ``L-blocks'' (for example, the number ``1'' and ``7'', ``0'' and ``8''). We choose parameters of error correcting code that is able to tolerate $4$ bits noise for every $63$ bits. Note that modifying a ``L-blocks'' may result in two bits flipped, thus, the probability that an attacker can modify the visual cue of a barcode  to another is less than  $\mathrm{\Phi}(3; 14, 0.75) = 3.98\%$ where $\mathrm{\Phi}$ is the cumulative distribution function of the binomial distribution $\mathbf{B}(14, 0.75)$.

\subsubsection*{Modifying control points} \ \ \
The adversary may try to modify the control points and this may cause failure in decoding, giving a string of  ``random'' bits which is unlikely to pass the MAC authentication check. Hence,  modifications of control points at most amount to a denial of service attack, which is not our main concern.

\section{Visual Cues for Verification of Multiple Barcodes}
\label{sec:subregionauthentication}

In this section, we discuss a few designs of visual cue, in particular,  for barcodes appeared in a linear sequence, and barcodes rendered as table. Recall that the main purpose of the visual cue is to bind location information to the barcodes, so that {\tt User} can visually verified that the barcodes are in the correct arrangement.

\subsubsection*{Linear Sequential Barcodes.} \ \ \
Consider a sequence of  barcodes  appearing  in the  order $B_1, B_2, \ldots, B_n$. The order of appearance gives implicit  structure of the encoded message. For instance, the message could be a string divided into substrings where each substring is encoded in a single barcode.  Hence, it is important to protect the order of appearance, even if the user may not interested in viewing all of them. A natural visual cue would be a counter, starting from 1, that is, the visual cue of block $B_i$ is $i$.  To indicate the end of the sequence, the last block contains a special symbol, say ``.'' in our example, to indicate end of sequence.

\subsubsection*{Barcodes in Table Structure.} \ \ \
Consider  a table of messages where each message is encoded in a barcode. The barcodes are  depicted in the natural table arrangement: for any 2 messages in the same row, the corresponding   barcodes are also in the same row, and  likewise for  columns. To protect such correspondence, we propose the following rules of assigning the visual cue: \\
\hrule
\begin{enumerate}
\item[R1] The numerical value of the visual cue symbol on the top row, leftmost block is 1. The value increments by 1 from left to right. At the end of the row, the increment process continues at the leftmost block of the row below if any.
 \item[R2] The rightmost block in each row has the additional cue which is a black dot indicating this is the end of row.
\item[R3] The rightmost block in the bottom row has an additional large black rectangle indicating this is the last block.
\end{enumerate}
\hrule

\ \\

Figure~\ref{fig:OurScheme:MS1} shows an example of such barcode table.
To verify that a table of barcodes are in the correct arrangement, {\tt User} simply needs to verify the continuity of the counter, every but the last row ends with a small dot, and the last barcode ends with a big dot. It is easy to verify that by imposing the above rules, any insertion, deletion or rearrangement of the barcodes can be detected by visual inspection.

\section{Alternative Methods}
\label{sec:alternative}

\begin{figure}[!t] \centering
\subfigure[{\tt Mobile} captures every blocks, then verifies and renders the whole message.]{\includegraphics[width=0.38\textwidth]{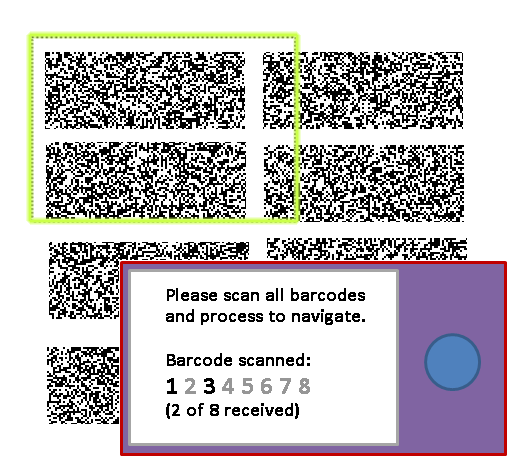}
\label{fig:alternatives:first}}
\subfigure[{\tt Mobile} displays the location(row/column) information encoded in the barcodes.]{\includegraphics[width=0.38\textwidth]{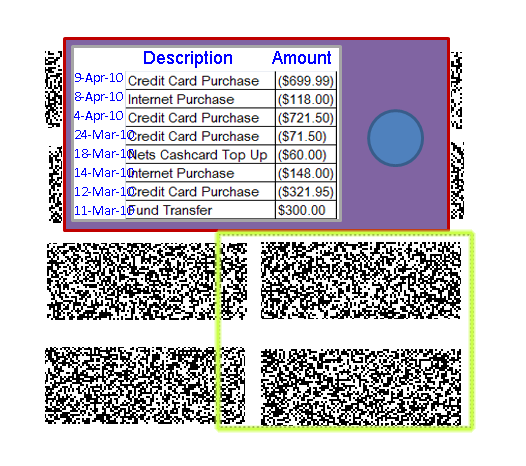}
\label{fig:alternatives:third}}
\caption{Illustration of alternative methods (for simplicity, only the barcodes and mobile device are shown here).}
\label{fig:alternatives}
\end{figure}

Besides using visual cues, there are other techniques to ensure that the barcodes are in correct order. This section compares our scheme with a few  alternatives. In general, our scheme uses more pixels to carry the visual cue symbols. On the other hand, it has the following advantage: (1) It does not disrupt the user by requiring the user to scan all the barcodes.  (2) It does not require the user to count the blocks on the terminal's display unit to verify the current block sequence on the mobile. (3) It allows the placement of barcodes to spread across different positions in a scrolling page, or even in different pages.
A brief illustration of the alternative methods is given in Figure~\ref{fig:alternatives}.

\subsubsection*{Embedding a HMAC of all blocks.} \ \ \ In this method,
given a long message $m_{\tt S}$, {\tt Server} computes a HMAC for the whole $m_{\tt S}$ and embeds $m_{\tt S}$ and its tag into a few barcodes. During authentication, the user first scans across all the barcodes, then {\tt Mobile} responds whether the HMAC agree with the content in the barcodes (Figure~\ref{fig:alternatives:first}). If so, {\tt Mobile} renders the long message and user navigates to obtain the required information. The advantages of this method are (1) the user does not need to verify the visual cue, and (2) the barcode is more efficient in the sense that it does not need to
embed the visual cue.

However, there are a few disadvantages of this method. Firstly, the scanning process could be less preferred when the user only want to browse a subset of the message (e.g. a user who wants to check a particular record from a list of transactions).  Secondly,
it is not easy to
navigate using the
relatively smaller
display panel in the mobile device.
 Furthermore,  it is not clear how to extend this method to the models where {\tt Mobile} device is not trusted.

\subsubsection*{Encoding location hints in barcode.} \ \ \
When the message can be represented as a form of table, one may try to secure the authenticity by using the row and column attributes as location information: Given a table $m_{\tt S}$, {\tt Server} first divides it into sub-tables, then it encodes each sub-table together with the corresponding row and column attributes into barcodes. When {\tt Mobile} decodes the barcode, it shows the corresponding attributes of the sub-table as shown in Figure~\ref{fig:alternatives:third}.

The advantage of this method is that it does not require the user to scan barcodes or verify visual cues, and the user can readily browse a sub-table of interest. While rearrangement attack can be prevented as the row and column information are encoded in the barcode, this method subjects to deletion attacks: the adversary may remove or duplicate an entire row of barcode without being detected. Although this could be patched by encoding more information (e.g. the total number of barcodes), the verification cost will increase (the user needs to count the barcode blocks).

\section{Implementation}
\label{sec:implementation}

The useability of our proposed method can be improved using ``augmented reality'' as described in the introduction.  We implemented a proof-of-concept system using webcam and laptop to simulate the mobile device.

\subsubsection*{Deploying Machines and Softwares.}\ \ \
We simulate the mobile
device and its camera using a Thinkpad X200 notebook (Intel core 2 duo $2.26$GHz CPU and $2$GB memory) equipped with an inexpensive usb webcam. To simulate the computing power of a typical mobile device, we allocate only $10\%$ CPU time and $128$M memory for our program. We use a Dell desktop machine with Intel Core 2 Duo $2.33$GHz CPU with $4$GB
of memory and Windows XP SP3 to simulate the network terminal.
The resolution of the webcam is $640 \times 480$ pixels with a maximum frame rate of $30$ frame per
second.
We tested the system on three
different display units: (1) a 19 inch flat TFT monitor in Dell model Optiplex
755; (2) a 15 inch flat TFT Dell UltraSharp monitor; and (3)
a 15 inch Dell CRT monitor. All
configuration of the display units such as brightness resolution are
reset to the default setting. In the following sections, we
call these three display units monitor 1, monitor 2 and monitor 3
respectively.

We use OpenCV libraries~\cite{opencvWeb} for basic image processing operation and
interfaces to the camera.

\subsubsection*{Choice of Parameters.}\ \ \
We use AES with 128 bit key for encryption scheme, HMAC based on SHA1 for message authentication code,   and
calculator fonts of numeric digits as visual cues symbols.
We use a $(63,36,11)$-BCH error correcting code~\cite{hocquenghem1959codes,bose1960class} to correct errors. That is, for every $36$ bits, we add $27$ bit of redundancy  and we are able to correct $5$ error bits. However, to prevent modification of visual cue, we reject to decode if there are more than $3$ error bits.

\subsubsection*{Image Processing Issues.}\ \ \
We use oversampling technique to reduce the noise of a captured image: one bit in the barcode is rendered using $2\times 2$ pixels. Let us call a group of $2\times 2$ pixels a ``superpixel''. Such oversampling can reduce the noise due to
mis-alignment and mitigating other artifacts, but it also reduce the channel capacity by a factor of $4$.

We use landmark-mapping~\cite{brown1992survey} method for image registration. That is, after {\tt Server} generates the barcodes, it super-imposes a set of 2D points $P$ called {\em control points}, whose position is known by {\tt Mobile}, on the barcode image.

After {\tt Mobile} captured a screenshot, it extracts the control points and find the best geometric transformation that maps the extracted control points to their original locations.
In our implementation, we find the best linear transformation that matches the points. The transformation is then applied to the barcode image.

\section{Performance}
\label{sec:performance}

In this section we measure the performance of our proof-of-concept implementation in terms of error rate, frame rate and channel capacity.

\subsubsection*{Image Registration Error.}\ \ \
To measure the accuracy of our image registration, we first generate an image of many blue points with the red control points.
The image is displayed on the three display units and captured by the camera. Image registration is then carried out and the displacement of blue points are measured. Here we use the Euclidean distance to measure the amount of displacement.

\begin{figure}[!t] \centering
\subfigure[Histogram of the displacement.]{\includegraphics[width=0.47\textwidth]{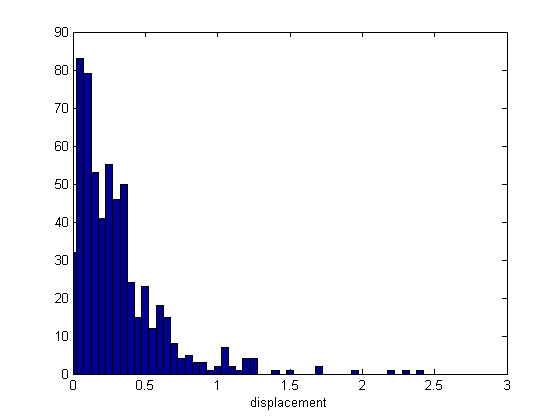}
\label{fig:histogramofposoff}}
\subfigure[The error rate of the three monitors over different frames.]{\includegraphics[width=0.43\textwidth]{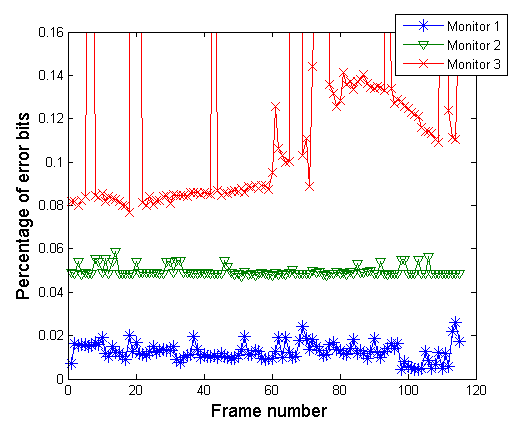}
\label{fig:errorrate}}
\caption{The performance of implementation.}
\label{fig:histogramanderrorrate}
\end{figure}

Our camera is able to capture a region of around $20$ blue points. Figure~\ref{fig:histogramofposoff} shows the histogram of the displacement
of all the blue points on monitor 1. Note that the average displacement is less
than $1$ pixel. The image registration algorithm can be
further refined by incorporating more effective and efficient known techniques.

\subsubsection*{Noise Level in Capturing Superpixels.}\ \ \
We now measure the probability of error in reading a superpixel. The camera is able to capture a block that consists of around $200
\times 150$ superpixels at a time.

After registration, we count the mismatches of superpixel between the registered image and the original image.
$30$ measurements are taken for each of the three display units. Figure~\ref{fig:errorrate} shows the
error for each measurement.

\subsubsection*{Frame Rate.}\ \ \
 The frame rate of our implementation is over $15$ frames per second running on the laptop machine. Although the implementation is not tested on mobile device, we believe a typical mobile device that has similar processing power could achieve more than $10$ frames per second, which is acceptable for most applications.

\subsubsection*{Capacity of Visual Channel.}\ \ \
We now give calculation for the size of payload (size of $m_{\tt S}$, the message {\tt Server} sends to {\tt User}) that can be embedded in a block that
occupies $10000$ pixels of {\tt Terminal}'s display unit. Recall that we used
$2 \times 2$ pixels to encode $1$ bit of the barcode, employed a $(63,36,11)$
BCH error correcting code, and used L-block to preserve the related
location. Thus the payload is $10000 \times \frac{1}{4} \times
\frac{36}{63} \times \frac{2}{3} = 952$ bits for such a block.

\section{Related Work}
\label{sec:relatedwork}

There is an extensive amount of
literatures exploiting the camera as an additional {\em visual
channel} for communication. Jacobs et al.~\cite{jacobs1996method}
gave a method that establishes a channel from a controllable light
source to a camera. McCune et al. proposed seeing-is-believing~\cite{mccune2009seeing}, which carries out authentication and
key-exchange over a visual channel established between a device's
display and another device's camera. Wong et al.~\cite{wong2007multi} built a prototype on a Nokia Series 60
handphone that provides 46 bits for authentication over the visual
channel.

Data can be transmitted to a camera effectively using 2D barcodes.
There are many 2D barcode designs, for example,  QR
code~\cite{code2000international} and the High Capacity Color
Barcode (HCCB)~\cite{parikh2008localization} that uses colored
triangles. Many barcodes
are designed to encode data in  printed copies. There are also
proposals that use other types of sources in the visual channel. Collomosse
et al. proposed ``Screen codes''~\cite{collomosse2008screen} for
transferring data from a display to a camera-equipped mobile device,
where the data  are encoded as a grid of luminosity fluctuation
within an arbitrary image.  A challenging hurdle in using hand-held
cameras to establish the channel is motion blur. A few stabilization
algorithms are developed for handheld
camera~\cite{sorel2005blind,or2007hand}, and for  2D barcodes~\cite{chu2007image}.

Similar to our scheme, Costanza et al.~\cite{costanza2009designable} suggested a technique to embed designs into barcodes to increase the expressiveness and to bring visually meaning to them. These systems recognize the barcodes based on the topology, rather than geometry, of the codes~\cite{costanza2003region}, and were initially developed for tracking objects in tangible user
interfaces and augmented reality applications~\cite{costanza2003introducing}.
Augmented reality has been exploited to enhance user experience on many applications including education~\cite{klopfer2008environmental}, gaming~\cite{squire2007mad},
outdoor activities~\cite{takacs2008outdoors}. Rekimoto et al.~\cite{rekimoto2000cybercode} Using 2D barcodes as the visual tags in
the augmented reality environment, where a camera can capture the barcode on physical object and link them to their information.

\section{Conclusion}
\label{sec:conclusion}
In this paper, we investigated how visual channel can be deployed to enhance security of the communication between server and user in various settings.  We pointed out that although authentication of an individual barcode can be easily carried out, the interesting technical challenge is in the verification of the relationships among several barcodes. This leads us to look into the problem of ``subregion authentication'' where a user wants to verify selective small pieces of data within a large dataset.  Although there are a few methods to overcome the problem, they introduce disruptions during the interactive session and are thus less user-friendly. To achieve seamless interactions, we proposed using visual cue to bind location information to the barcode, so as to aid the user in visually verifying the data.

Our protocols demonstrated that, the visual channel ``enhanced'' with the visual cue, together with the mobile device's input/output device, jointly provide more flexibility in designing secure protocols. Viewing from another perspective, our investigation highlights limitations of visual channel, for instance, the observation that  confidentiality is difficult to achieve under the setting where either the mobile device or the terminal could be dishonest.
Our solution serves as an interesting example where security is achieved by coupling computer's processing power with human perceptual system. The design of our barcode also serves as an interesting application of fragile watermark.

To demonstrate the concept, we give a system that simulates the
mobile device using webcam and laptop. The performance of the system is
promising. Although we have not yet implemented
the framework on actual mobile device, we believe that the
processing power of many current mobile devices is sufficient to
provide seamless interactions.

\bibliographystyle{plain}
\bibliography{mob-arxiv}

\end{document}